\documentclass[amsmath,amssymb,superscriptaddress,nobalancelastpage,prl,twocolumn,longbibliography]{revtex4-1}

\usepackage{graphicx}
\usepackage{varioref}
\usepackage{xr-hyper}
\usepackage{xcolor}
\usepackage{nicefrac}
\usepackage{xfrac}
\usepackage{hyperref}
\hypersetup{colorlinks,linkcolor=blue,urlcolor=blue,citecolor=orange}
\usepackage{ulem}
\usepackage{siunitx}
\usepackage{graphicx}% Include figure files
\usepackage{dcolumn}% Align table columns on decimal point
\usepackage{bm}% bold math
\usepackage{braket}

\begin{document}

\title{Spatially Modulated Superfluid State in Two-dimensional $^4$He Films}

\author{Jaewon Choi}
%\thanks{These authors contribute equally to this work.}
\altaffiliation{Current affiliation: Diamond Light Source, Ltd., Didcot, Oxfordshire OX11 0DE, United Kingdom}
\affiliation{Department of Physics, Korea Advanced Institute of Science and Technology (KAIST), 291 Daehak-ro, Yuseong-gu, Daejeon 34141, Republic of Korea}

\author{Alexey A. Zadorozhko}
%\thanks{These authors contribute equally to this work.}
\altaffiliation{Current affiliation: Quantum Dynamics Unit, Okinawa Institute of Science and Technology (OIST), 1919-1 Tancha, Okinawa 904-0495, Japan}
\affiliation{Department of Physics, Korea Advanced Institute of Science and Technology (KAIST), 291 Daehak-ro, Yuseong-gu, Daejeon 34141, Republic of Korea}

\author{Jeakyung Choi}
\affiliation{Department of Physics, Korea Advanced Institute of Science and Technology (KAIST), 291 Daehak-ro, Yuseong-gu, Daejeon 34141, Republic of Korea}

\author{Eunseong Kim}\email{eunseong@kaist.edu}
\affiliation{Department of Physics, Korea Advanced Institute of Science and Technology (KAIST), 291 Daehak-ro, Yuseong-gu, Daejeon 34141, Republic of Korea}

\begin{abstract}
The second layer of $^4$He films adsorbed on a graphite substrate is an excellent experimental platform to study the interplay between superfluid and structural orders. Here, we report a rigid two-frequency torsional oscillator study on the second layer as a function of temperature and $^4$He atomic density. We find that the superfluid density is independent of frequency, which can be interpreted as unequivocal evidence of genuine superfluidity. The phase diagram established in this work reveals that a superfluid phase coexists with hexatic density-wave correlation and a registered solid phase. This suggests the second layer as a candidate for hosting two exotic quantum ground states: the spatially modulated superfluid and supersolid phases, resulting from the interplay between superfluid and structural orders.
\end{abstract}

\pacs{74.72.-h, 71.45.Lr, 74.25.Dw}

\maketitle

Interplay among different orders is a hallmark of correlated quantum systems \cite{Fradkin2015,Davis2013,Fernandez19}. High-temperature superconductors \cite{Keimer2015,AgterbergAnnRev2020,Chang12}, heavy-fermion compounds \cite{Gerber2014,DYKim2016}, and low-dimensional materials \cite{Haddad2002,Zhai2013,Ugeda2016} have been studied as model systems, where multiple broken-symmetry phases are intertwined and exotic quantum phases emerge. Similarly, each layer of $^4$He films adsorbed on graphite substrates allows for the realization of exotic quantum phenomena in a two-dimensional system, subject to periodic triangular potential. For example, the first layer undergoes successive structural transitions due to strong substrate potential \cite{Corboz2008,Gordillo2009}. Superfluidity found in the third layer is an experimental manifestation of the Berezinskii-Kosterlitz-Thouless (BKT) phase transition \cite{Berezinskii1972,Kosterlitz1973,Crowell1996}.

In the second layer, a delicate balance between $^4$He-$^4$He and $^4$He-graphite interactions plays a significant role. It has accordingly been widely studied as a candidate for hosting a supersolid ground state in which both long-ranged superfluid and solid orders coexist in a single homogeneous phase. Indeed, superfluidity was reported in torsional oscillator (TO) experiments \cite{Crowell1993,Crowell1996,Nyeki2017} near the $^4$He atomic density (or coverage) where the signature of a solid melting in the heat capacity was observed \cite{Greywall1991,Greywall1993,Nakamura2016} and the existence of a commensurate solid phase has been predicted \cite{Pierce1998,Ahn2016,Gordillo2020}. The unusual log($T$) behavior and the absence of a sharp onset in the previous TO experiments cannot be understood in the framework of BKT theory alone. Recently, it was suggested that the intertwined superfluid and density-wave orders lead to a sixfold softening of the rotonlike mode, which explains the lack of BKT transition \cite{Nyeki2017,Lieu2019}. Thus, the strong interplay between the superfluid and structural orders may be the key to understanding the exotic superfluid behaviors and exploring the possible supersolid phase in the second layer.

\begin{figure*}[t]
\center{\includegraphics[width=0.9\textwidth]{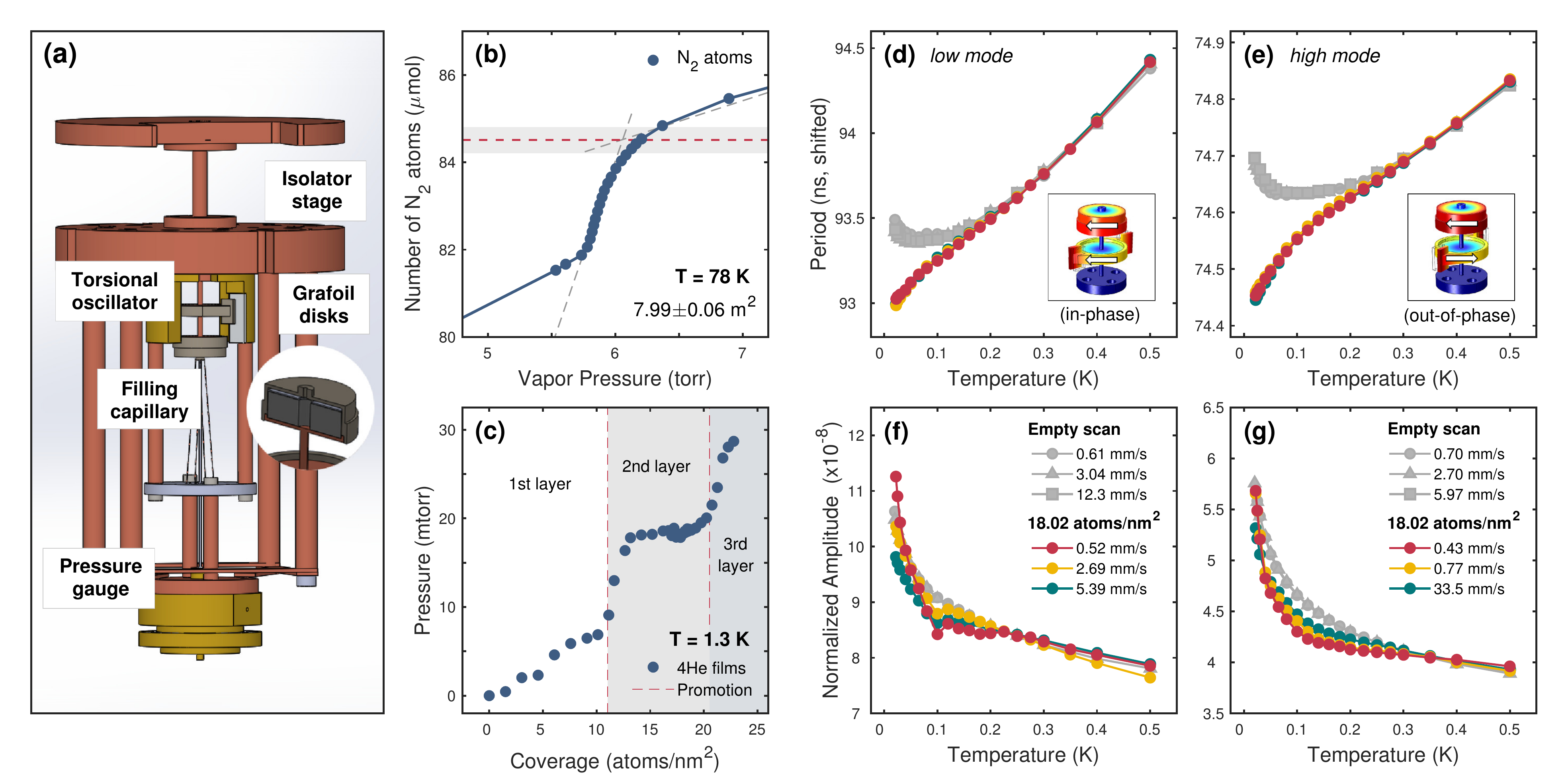}}
\caption{\textbf{Superfluid measurement in two-dimensional $^{4}$He films.} (a) Rigid two-frequency torsional oscillator (TO). (b) N$_{2}$ pressure isotherm for the determination of the surface area of substrate. (c) Layer-by-layer growth of $^4$He films manifested by discontinuous increases in vapor pressure. (d), (e) Temperature dependence of the TO period, and (f), (g) amplitude in the low (in-phase) and high (out-of-phase) modes at different rim velocities. Empty TO responses (gray symbols) obtained at different driving velocities are superimposed over each other, indicating that the TO operates in the linear response regime. 
}	
\label{fig1}
\end{figure*}

However, two outstanding questions still remain to be addressed: First, can we attribute the TO responses to the emergence of superfluidity? Since the oscillating motion of TOs can also be influenced by mechanisms other than superfluidity \cite{Reppy2012}, experiments that disentangle the superfluid contribution from that of nonsuperfluid origins are necessary. Second, do the superfluid and structural orders coexist in a common coverage range or lie in separate coverage regions? It would be interesting to understand their relationship, if they indeed coexist.

Here, we present a two-frequency TO study on the second layer of $^4$He films adsorbed on graphite to address the above-mentioned questions. The superfluid density $\rho_{s}$ is measured as a function of temperature and coverage. $\rho_{s}$ measured at two different frequencies are equivalent, indicating that the TO responses can be credited to genuine superfluid transition. By \textit{in situ} pressure measurement, the phase diagram for the second layer is more accurately determined and compared to other studies. The superfluid phase emerges in the liquid region where the superfluid order with hexatic density correlation was predicted \cite{Gordillo2020}. The superfluid order is rapidly suppressed in the region where both superfluid and registered solid phase coexist, revealing a competing relationship between them. This coexistence region can be interpreted as a candidate for the long-sought supersolid phase.

\begin{figure}[t]
\center{\includegraphics[width=0.47\textwidth]{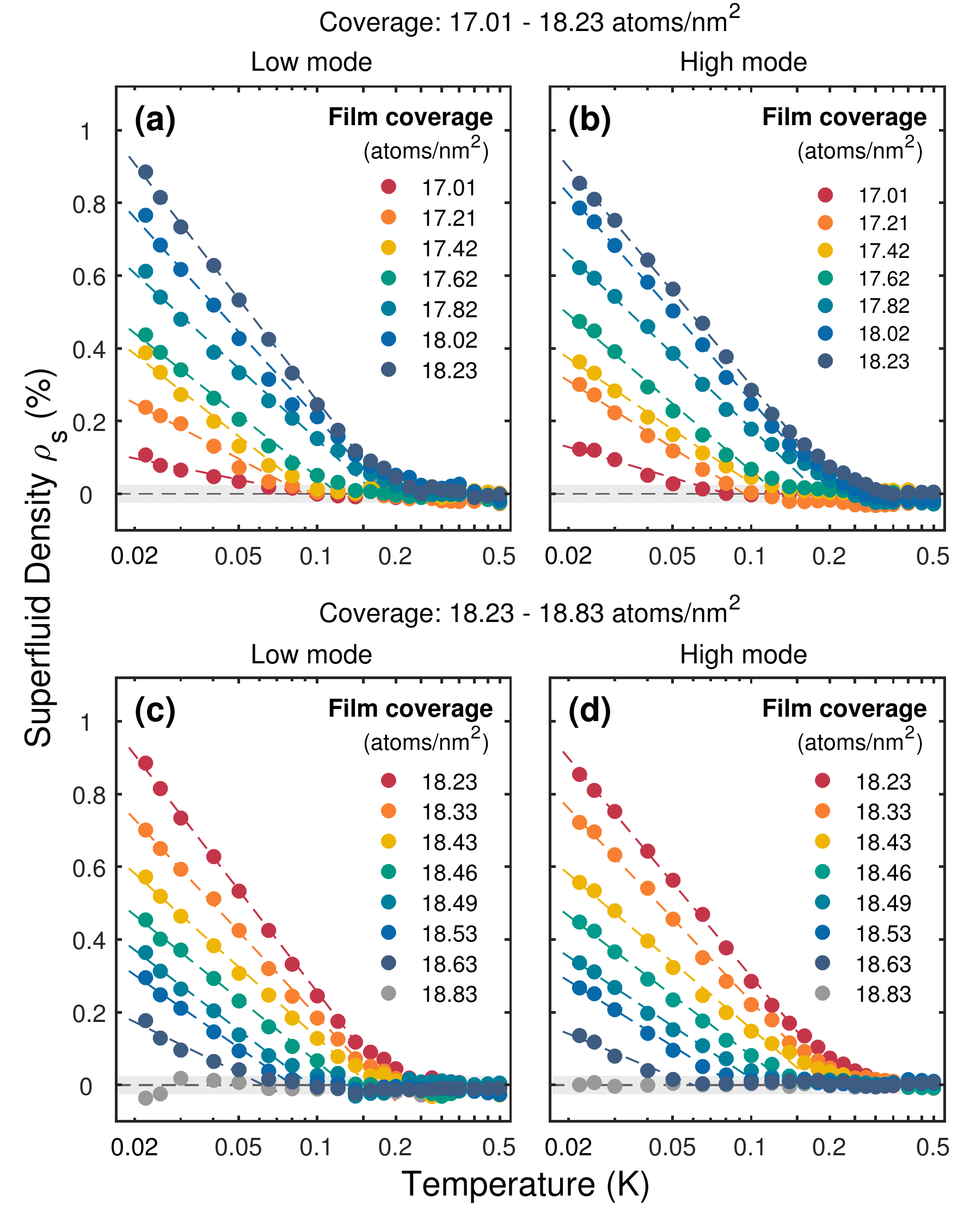}}
\caption{\textbf{Temperature evolution of second layer superfluidity.} Superfluid density $\rho_{s}$ as a function of temperature $T$ in a coverage region of 17.01$\sim$18.23 atoms/nm$^2$ for the (a) low- and (b) high-frequency modes, and 18.23$\sim$18.83 atoms/nm$^2$ for the (c) low- and (d) high-frequency modes. The dashed lines are linear fits of $\rho_{s}$ as a function of log($T$).
}	
\label{fig2}
\end{figure} 

For unambiguous detection of the superfluid phase, a rigid two-frequency TO containing a Grafoil substrate was fabricated [Fig.~\ref{fig1}(a)]. Unlike conventional single-mode TOs \cite{Crowell1993,Crowell1996,Nyeki2017}, it measures $\rho_{s}$ at two different frequencies: 511 Hz (low mode, $f_{-}$) and 1246 Hz (high mode, $f_{+}$), with a quality factor of $10^6$. An \textit{in situ} diaphragm-type pressure gauge was installed to measure the vapor pressure of the $^4$He films. Figure~\ref{fig1}(b) shows the N$_2$ pressure isotherm of the Grafoil substrate determining its surface area. Because of the completion of the $\sqrt{3}$ $\times \sqrt{3}$ commensurate solid phase at 6.37 atoms/nm$^2$ \cite{Chan1984}, the slope changes suddenly at $8.45\times10^{-5}$ mol corresponding to $7.99\pm0.06$ m$^2$. Ultrapure $^4$He gas with a $^3$He impurity concentration of 0.6 ppb was systematically dosed to the sample cell and annealed at high temperatures. We confirm the layer-by-layer growth \cite{Zimmerli1992,Noury2019} via the $^4$He vapor pressure isotherm [Fig.~\ref{fig1}(c)] that shows two clear jumps at 11.1 and 20.4 atoms/nm$^2$ (see Fig.~1 in the Supplemental Material).

Figures~\ref{fig1}(d)-\ref{fig1}(g) show the typical temperature dependence of the TO period and amplitude, measured at a $^4$He atomic density $n$ (or coverage) of 18 atoms/nm$^2$. In both modes, the period deviates from that of the empty TO at the onset temperature $T_{s}$ $\sim$250 mK upon cooling [Figs.~\ref{fig1}(d) and \ref{fig1}(e)]. Our TO was specifically designed to distinguish the physical origin of this period reduction, $\delta P_{\pm}$. If part of the $^4$He films undergoes a superfluid transition below $T_c$, it decouples from the oscillation of the TO and hence does not contribute to the rotational inertia of $^4$He. This ``missing'' inertia leads to a decrease in the resonant period of the TO, $\delta P_{\pm}$. The superfluid density $\rho_{s}(T,n)$ is then determined independent of frequency by $\rho_{s}=\delta P_{-}(T,n)/\Delta P_{-}(n)=\delta P_{+}(T,n)/\Delta P_{+}(n)$, where $\Delta P_{\pm}(n)$ is the period increase at 500 mK due to the $^4$He mass added to the second layer. Other mechanisms such as viscoelastic property change \cite{Day2007,Reppy2012} or slippage of $^4$He atoms on the substrate \cite{Hieda2000,Hosomi2008}, on the other hand, produce nontrivial frequency responses. For example, the viscoelastic stiffening of solid $^4$He induces a superfluid-mimicking period reduction $\delta P/\Delta P$ proportional to ${f}^2$, virtually indistinguishable from genuine superfluid transition by single-mode TOs \cite{Reppy2012}. The TO amplitude shows a broad dip in the temperature range where the period reduction appears [Figs.~\ref{fig1}(f) and \ref{fig1}(g)]. The period curve remains unaffected upon increased oscillation speeds; thus, the superfluid critical velocity $v_{c}$ exceeds 33 mm/s, consistent with that of $^4$He films \cite{Chan1988}.

Figure~\ref{fig2} shows $\rho_{s}(T)$ measured at various coverages in the second layer. We first observe $\rho_{s}$ above our detection limit at $n_c=17$ atoms/nm$^2$. A TO period and amplitude below $n_{c}$ are superimposed with a vertically shifted empty-TO background (see Fig.~2 in the Supplemental Material). The absence of a ``tilted" or ``composite" background \cite{Choi2015,Choi2018} confirms that our TO adopting rigid design principle is unaffected by the complicated viscoelastic coupling between the $^4$He films and TO body,  observed in other nonrigid TOs \cite{Crowell1993,Crowell1996,Nyeki2017}. As $n$ increases to 18.23 atoms/nm$^2$, $\rho_{s}$ reaches its maximum value of $\sim$0.9 \%. Above this coverage, $\rho_{s}$ is rapidly suppressed, and disappears at 18.83 atoms/nm$^2$. Remarkably, the $\rho_{s}$ measured at two different frequencies are essentially identical. This observation of a frequency-independent $\rho_{s}$ can be interpreted as distinctive evidence of genuine superfluidity in the second layer. This conclusion is supported by TO responses of BKT superfluid in the third layer (see Fig.~3 in the Supplemental Material) \cite{Crowell1993,Crowell1996}. We also find $\rho_{s}$ to be frequency-independent in the third layer, confirming our rationale.

\begin{figure*}[t]
\center{\includegraphics[width=0.95\textwidth]{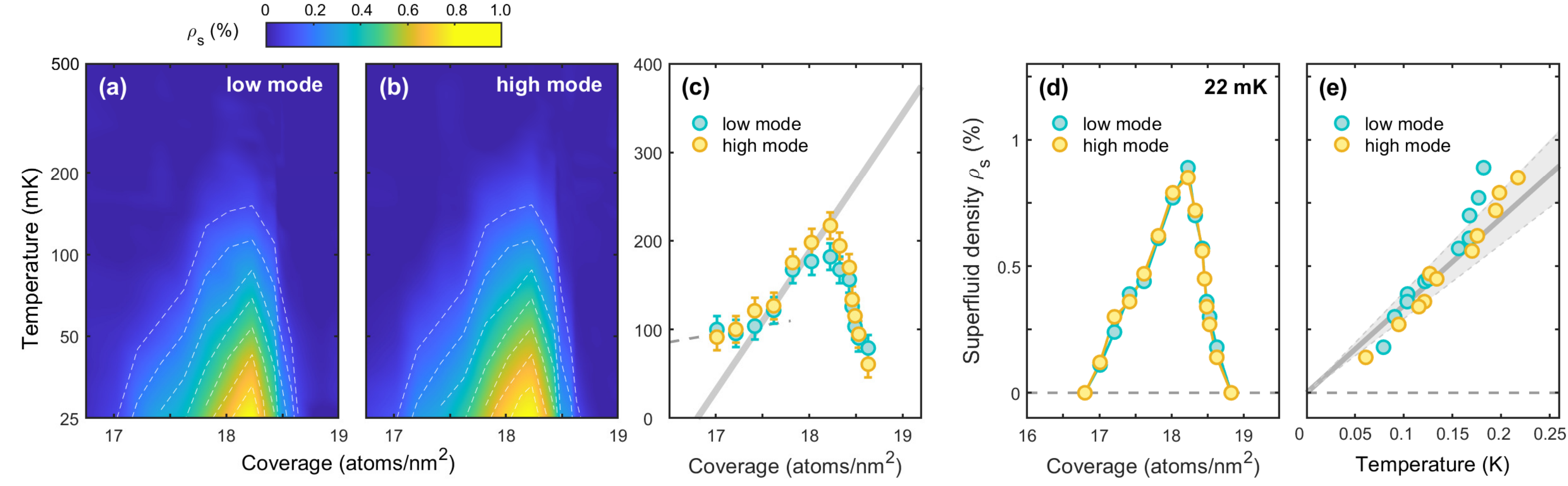}}
\caption{\textbf{Coverage dependence of the second layer superfluidity.} False-color map of superfluid density $\rho_{s}$ in the second layer as a function of coverage $n$ and temperature $T$ measured at (a) the in-phase and (b) the out-of-phase modes. The dashed lines are contours of $\rho_{s}$ from 0.1 (outermost) to 0.7 (innermost) with a step of 0.1. (c) Superfluid onset temperature $T_{s}$ as a function of $n$. The gray solid line is the prediction from BKT theory $T_{c}(n)=0.15(n-n_0)$ \cite{Bishop1980}. (d) $\rho_{s}$ isotherm at 22 mK plotted against $n$. (e) Testing the linear relation between $\rho_{s}$ and $T_{s}$. The solid gray line is a linear fit fixing the y intercept to zero. The shaded region indicates 1$\sigma$ bound (or 68\% confidence) of the fitting.
}	
\label{fig3}
\end{figure*}

Besides the frequency-independent $\rho_{s}$, there are a few more remarkable features of the second layer superfluidity. First, the temperature evolution of $\rho_{s}$ is much slower than that expected from the BKT theory. This theory predicts a sudden increase in $\rho_{s}(T)$ at the onset temperature due to vortext-antivortex unbinding, followed by shallow phononlike excitation with $T^3$ dependence at low temperature \cite{Agnolet1989}. However, $\rho_{s}(T)$ measured in the second layer increases proportionally to log($T$) in both frequencies (Fig.~\ref{fig2}). This trend is in stark contrast to the superfluid behavior in the third layer, where a typical broadened BKT transition was observed (see Fig.~3 in the Supplemental Material). This $\rho_{s}\sim$ log($T$) with the slow onset has also been reported in previous TO experiments \cite{Crowell1993,Crowell1996,Nyeki2017}. Such unusual behavior was recently attributed to intertwined superfluid and density-wave orders and has been associated with a softening of the rotonlike modes \cite{Nyeki2017} or a ``failed" superfluid in absence of topologically stable vortices due to spatial modulation \cite{Lieu2019}. Second, $\rho_{s}$ does not strongly depend on the $^3$He impurity concentration. Although we used ultrahigh-purity $^4$He gas with 0.6 ppb $^3$He impurities several orders of magnitude smaller than other studies \cite{Crowell1993,Crowell1996}, similar $\rho_{s}$ and $T_{s}$ are observed. Therefore, $^3$He impurity does not lead to significant effects on the superfluid transition in the second layer. It is notable that the robust superfluidity at (or immunity to) the extremely low concentration of $^3$He is a typical characteristic of a BKT superfluid. Third, $\rho_{s}\sim$ 0.9\% can be understood by the tortuosity effect with $\chi\sim0.98$, measured near the third-layer completion. This indicates that a significant fraction of $^4$He atoms in the second layer participate in the superflow. Although $\rho_{s}(T)$ resembles the slow onset reported in the TO response in bulk solid $^4$He, the present observations of frequency independence, weak $^3$He dependence, and high critical velocity are interpreted as unequivocal evidence of superfluidity in the second layer.

%Second, superfluid energy dissipation $\Delta Q^{-1}$ is vastly suppressed. As seen in Fig.~\ref{fig2}, $\Delta Q^{-1}$ is too small $\sim10^{-8}$ or too broad to be clearly observed, in stark contrast to evident dissipation in the third layer (Fig.~\ref{fig3}H and Supplementary Fig. 3). Superfluid transition in two dimension accompanies with energy dissipation associated with diffusive motion of 2D vortices driven by oscillating superflow \cite{Huberman1978,Ambegaokar1978}. This is uniquely observed in 2D superfluid driven by KT phase transition \cite{Bishop1978,Bishop1980}, but not in bulk superfluid or in percolating superfluid model \cite{Taborek2020}. Thus, the absence or suppression of $\Delta Q^{-1}$ implies the 2$^\textrm{nd}$-layer superfluidity cannot be understood in the framework of KT phase transition theory.

In Figs.~\ref{fig3}(a) and \ref{fig3}(b), we map out $\rho_{s}(T,n)$ in the second layer at both frequencies. The white dashed lines are the contours of $\rho_{s}$. The maps reconfirm the frequency independence of the superfluid over the entire parameter space. The superfluid onset $T_{s}$ determined in Fig.~4 of the Supplemental Material is plotted in Fig.~\ref{fig3}(c) against $n$. Until $n$ reaches $\sim$17.6 atoms/nm$^2$, $T_{s}(n)$ increases slowly. Above this value, $T_{s}(n)$ grows much faster following a linear function of $n$, $T_{s}(n)\sim0.15(n-n_0)$ where $n_0=16.8$ atoms/nm$^2$ is the coverage of an inert layer. This slope is consistent with both third-layer superfluid and superfluid films adsorbed on mylar \cite{Bishop1980}, which are well understood by the BKT phase transition. The similarity between the second and third layers further suggests that the observed superfluid responses are associated with the same region of the Grafoil substrate. If they stem from different regions--for example, one in the crests between crystallites and the other on the crystalline surface, the slope of $T_{s}(n)$ should be different due to the different areal densities. In addition, the $\rho_{s}$ isotherm at 22 mK is shown in Fig.~\ref{fig3}(d). The BKT relation $\rho_{s}/T_{s}=8\pi k_{B}m^2/h^2$ predicts that $T_{s}$ is linearly proportional to $\rho_{s}$. Figure~\ref{fig3}(e) demonstrates that the second layer superfluid on graphite satisfies the linear relation, which has been found in $^4$He films on various substrates \cite{Csathy2003}. This implies that the superfluid phase in the second layer might not be entirely distinct from the BKT superfluid found above the third layer, although strong interplay with structural order suppresses its sharp onset.

\begin{figure}[t]
\center{\includegraphics[width=0.47\textwidth]{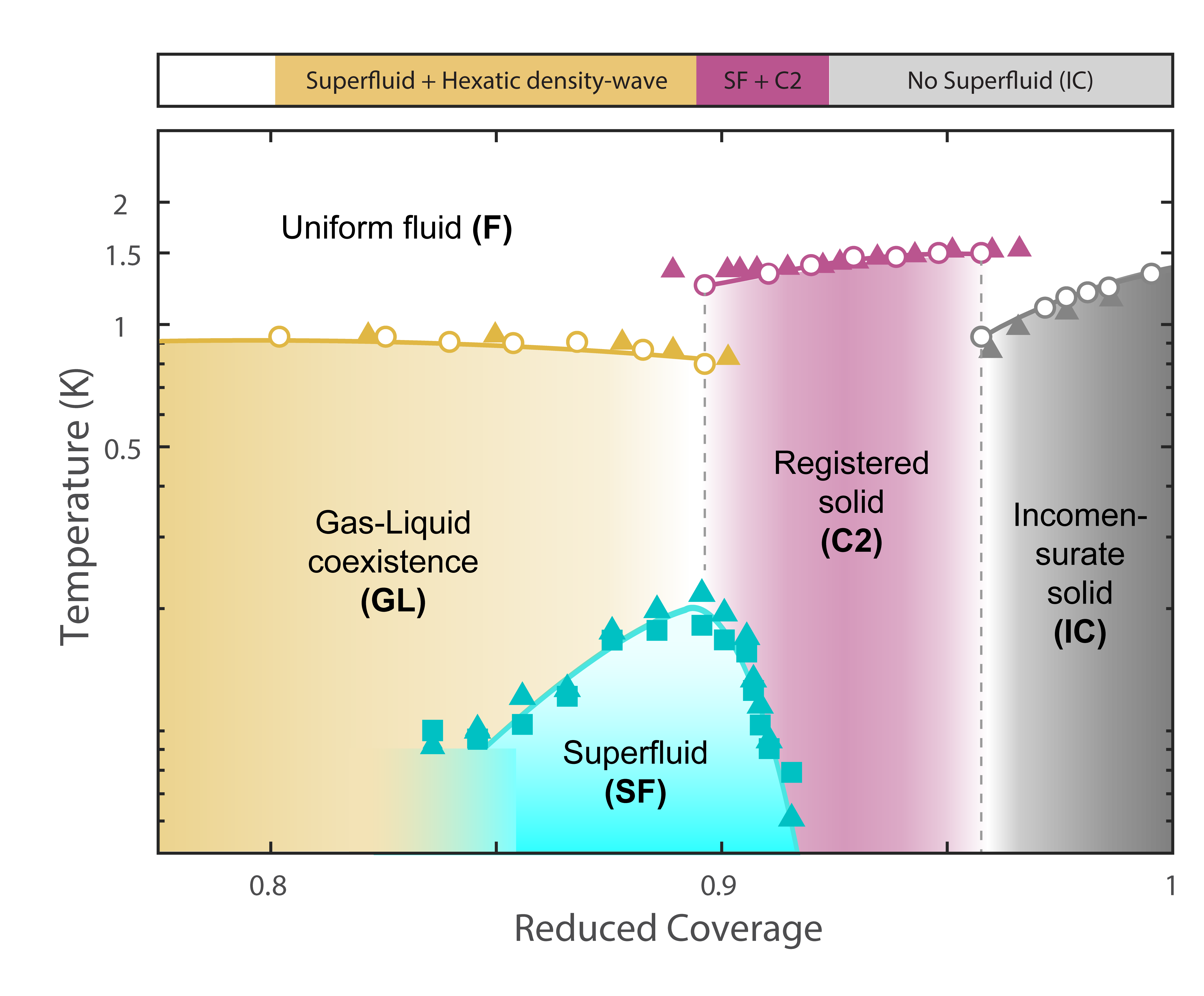}}
\caption{\textbf{Phase diagram of the second layer of $^4$He films.} $T_{s}$ from this work (cyan) and the temperature of heat capacity loci (open circles and solid triangles) are plotted against reduced coverage (see text) with $n_2=21.2$ atoms/nm$^2$ \cite{Greywall1993,Nakamura2016}. A theoretical phase diagram \cite{Gordillo2020} is drawn above the figure for comparison under the assumption of $n_2$ = 20.4 atoms/nm$^2$. The uncertainty of $n_r$ is smaller than 0.01. 
}	
\label{fig4}
\end{figure}

Simultaneous TO and \textit{in situ} vapor pressure measurements enable us to determine an accurate coverage scale. Based on this, we propose an $n$-$T$ phase diagram of the second layer in Fig.~\ref{fig4}. To reconcile with different experiments, we introduce a ``reduced" coverage $n_{r}=n/n_{2}$, where $n_{2}$ is the coverage for second layer completion. The superfluid onset $T_{s}(n)$ measured in this work is plotted with cyan symbols. Previously, three structural phases have been identified by heat capacity measurements \cite{Greywall1991,Greywall1993,Nakamura2016}; we incorporate the results applying our reduced coverage scale into the proposed phase diagram. The low-density region is assigned to a gas-liquid coexistence (GL) phase, evidenced by heat capacity peaks near 0.8 K nearly independent of $n$. At higher $n_\textrm{r}\sim0.9$, an additional heat capacity peak emerges near 1.5 K, which can be attributed to the appearance of a second layer commensurate solid phase (C2) analogous to $^3$He films on graphite \cite{Nakamura2016}. However, the existence and exact symmetry of the C2 phase are still questionable, as numerical studies do not converge into a single conclusion  \cite{Corboz2008,Ahn2016,Gordillo2020,Moroni2020}. Above $n_{r}\sim0.9$, the heat capacity peak is slowly replaced by a sharp peak near 1 K due to an incommensurate solid phase (IC). The superfluid phase coexists with the C2 phase in a narrow coverage region near $n_{r}\sim0.9$. We note that the uncertainty is $\delta n_{r}<0.01$, smaller than the range of coexistence.

%The onset temperatures of superfluid (SF) phase measured in this work are also presented in Fig.~\ref{fig4}. Above $n_\textrm{norm}\sim$ 0.76, both the onset temperature and $\rho_{s}$ are strongly suppressed. The point where the suppression starts rougly coincides with the coexistence region where the C2 phase starts to appear. In Fig.~\ref{fig4}, the phase diagram suggested by the most recent diffusion Monte Carlo (DMC) calculation study \cite{Gordillo2020}. All together, experimental results are consistent with the DMC calculation: The SF coexists with C2 phase in the second layer of $^4$He on graphite.

%The frequency-dependent study of superfluidity using a rigid TO provides unequivocal evidence of genuine superfluidity in the second layer, even though the slow onset behavior is reminiscent of viscoelastic anomaly found in bulk solid $^4$He \cite{Day2007}. In addition, we also confirm that both superfluid and density-wave order coexist in a narrow region of the second layer owing to simulatneous measurement of superfluid density and $^4$He vapor pressure and absoulte coverage scaling. The last outstanding question still remains to be addressed: What is the relationship between them?

The phase diagram proposed here is consistent with recent diffusive Monte Carlo (DMC) calculations \cite{Gordillo2020}. According to this work, the second layer undergoes a first-order phase transition from a low-density liquid into a 7/12 registered solid, stable within 18.2$\sim$18.6 atoms/nm$^2$. Above this coverage, an incommensurate triangular solid takes over near the layer completion. $\rho_s(T\rightarrow0)$ in the low-density liquid region was estimated to be nearly 1, whereas it had a partially suppressed value in the 7/12 registered solid. In this reference, the incommensurate solid phase does not support superfluidity. This result can explain the suppression of the superfluidity initiated near the structural transition as well as the absence of superfluidity in the incommensurate solid phase that is observed here. Furthermore, the DMC calculation found a hexatic density correlation induced by a corrugation of the first layer in the liquid region \cite{Gordillo2020}, implying that the superfluid order is spatially modulated.

The coexistence of superfluid and solid orders was also confirmed by the DMC calculation \cite{Gordillo2020}. $\rho_{s}$ is rapidly suppressed inside the C2 phase, indicating competition between two orders tuned by $n$. One question naturally arises here: How do they coexist in the second layer? The most intriguing answer is as a supersolid phase in which the superfluid and solid order coexist spontaneously in a single uniform state. Its possible observation was first reported by TO experiments with bulk solid $^4$He, but was later attributed to the viscoelastic property change \cite{Kim2004a,Kim2004b,Beamish2020}. The supersolid phase has also been intensively studied in spatially ordered dipolar gas \cite{Leonard2017a,Leonard2017b,Li2017}. In addition, recent observations of a softening of excitation spectra that lead to the development of roton minima have been interpreted as the signature of supersolidity \cite{Petter2019,Natale2019}. Although a perfect $^4$He single crystal might not be supersolid \cite{Prokofev2005,Boninsegni2006}, this phase has proven to exist in Bosonic triangular lattices \cite{Boninsegni2005,Wessel2005}. Otherwise, an alternative explanation for the coexistence is that certain region are phase-separated by forming domains with short-range density correlation. In this case, the superfluid occupies the area between the solid domains but its percolation gets weaker as more $^4$He atoms are added. If $n$ reaches some critical value where the registered solid does not support superfluid percolation, the superfluid disappears.

The spatially modulated superfluid phase observed here is a prime example of an exotic quantum phase resulting from the interplay among competing orders. Similarly, spatially modulated superconducting (SC) states have been widely explored in quantum materials where the $d$-wave SC state is coupled to the magnetic order \cite{Gerber2014,DYKim2016} and forms the Fulde-Ferrell-Larkin-Ovchinnikov state in high magnetic fields \cite{Matsuda2007}. In cuprate materials, the putative pair-density-wave (PDW) state, which is a spatially modulated SC order intertwined with spin- and charge-density-wave orders, has been suggested as a ``mother" state when an SC order strongly competes with them \cite{Fradkin2015}. Other descendant orders are ``secondary" orders, generated by sequential symmetry breaking of the PDW state. Despite possible signatures from recent STM works \cite{Du2020,Choubey2020}, the observation of this long-sought PDW state has been hindered by the intrinsic complexity of strongly correlated systems. The exotic superfluid phases observed here and the recent searches for the PDW state in $^3$He \cite{Levitin2019,Shook2020} imply that low-dimensional helium can provide excellent experimental platforms to realize novel quantum many-body phenomena for both bosons ($^4$He) and fermions ($^3$He), benefiting from their extremely pristine quality and reduced complexity \cite{Saunders20}.

In summary, our rigid two-frequency TO study of $^4$He films on graphite provides unequivocal evidence of the superfluid phase in the second layer, confined to a coverage range of 17$\sim$18.8 atoms/nm$^2$. $\rho_{s}(T,n)$ measured at two frequencies were almost identical, indicating the existence of genuine superfluidity. Based on an accurate coverage measurement, a refined phase diagram of the second layer was presented. The superfluidity first emerges in the low-density liquid where the previous DMC calculations reported a superfluid ground state with a hexatic density-wave correlation. The superfluid state persists up to a higher coverage where the registered solid phase emerges. The coexistence of superfluid and solid orders leads to strong suppression of $\rho_s$, revealing a competing relationship between them. This exotic superfluid state intertwined with a structural order can be understood with an analogy to the spatially modulated superconducting state predicted in quantum materials where interplay among various competing orders plays a central role. This result suggests two-dimensional $^4$He films as a model system for studying the interplay among broken-symmetry phases and for exploring exotic quantum ground states.

This work was supported by the National Research Foundation of Korea grant funded by the Korean government (MSIT) through MMO (2019R1A2C1009299) and the Center for Quantum Coherence in Condensed Matter (2016R1A5A1008184). J.C. especially thanks the POSCO TJ Park Foundation for its financial support through the TJ Park Science Fellowship. A. A. Z. and J. C. contributed equally to this work.

\bibliography{4He_ref}

\clearpage
\onecolumngrid

\newcommand{\beginsupplement}{
        \setcounter{table}{0}
        \renewcommand{\thetable}{S\arabic{table}}
        \setcounter{figure}{0}
        \renewcommand{\figurename}{\textbf{Supplementary Figure}}}

\beginsupplement
\clearpage
\onecolumngrid

\end{document}